\documentclass[preprint2]{aastex6}
\usepackage{amsfonts,amsmath,amssymb,ulem,nicefrac,verbatim,color}
\usepackage{array,graphicx}%longtable,ltcaption,lscape
\usepackage{affils}

\def\simlt{\mathrel{\hbox{\rlap{\hbox{\lower4pt\hbox{$\sim$}}}\hbox{$<$}}}}
\def\simgt{\mathrel{\hbox{\rlap{\hbox{\lower4pt\hbox{$\sim$}}}\hbox{$>$}}}}

\def\ale{\mathrel{\hbox{\rlap{\hbox{\lower4pt\hbox{$\sim$}}}\hbox{$<$}}}}
\def\age{\mathrel{\hbox{\rlap{\hbox{\lower4pt\hbox{$\sim$}}}\hbox{$>$}}}}

\def\spose#1{\hbox to 0pt{#1\hss}}

\begin{document}

%\title{Unusually Strong and High Frequency Diffractive Scintillation in GRB\,161219B}

\title{An Unexpectedly Small Emission Region Size Inferred from Strong High-Frequency Diffractive Scintillation in GRB\,161219B}

\DeclareAffil{cfa}{Harvard-Smithsonian Center for Astrophysics, 60 Garden St., Cambridge, MA 02138, USA}
\DeclareAffil{nrao}{National Radio Astronomy Observatory, 520 Edgemont Road, Charlottesville, VA 22903, USA}
\DeclareAffil{UCB}{Department of Astronomy, University of California, 501 Campbell Hall, Berkeley, CA 94720-3411, USA}
%\DeclareAffil{az}{Steward Observatory, University of Arizona, 933 N. Cherry Avenue, Tucson, AZ 85721, USA}
%\DeclareAffil{ef}{Einstein Fellow}
\DeclareAffil{bath}{Department of Physics, University of Bath, Claverton Down, Bath BA2 7AY, UK}
%\DeclareAffil{slov}{Faculty  of  Mathematics  and  Physics,  University  of  Ljubljana, Jadranska 19, 1000 Ljubljana, Slovenia}
\DeclareAffil{liver}{Astrophysics Research Institute, Liverpool John Moores University, IC2 building, Liverpool Science Park, 146 Brownlow Hill, Liverpool L3 5RF, UK}
\DeclareAffil{slov}{Center for Astrophysics and Cosmology, University of Nova Gorica, Vipavska 13, 5000 Nova Gorica, Slovenia}
\DeclareAffil{mex}{Instituto de Astronom\'{i}a, Universidad Nacional Aut\'{o}noma de M\'{e}xico, Apdo. Postal 70-264, Cd. Universitaria, DF 04510, M\'{e}xico}
\DeclareAffil{nu}{Center for Interdisciplinary Exploration and Research in Astrophysics (CIERA) and Department of Physics and Astronomy, Northwestern University, Evanston, IL 60208, USA}
\DeclareAffil{hubble}{Hubble Fellow}
\DeclareAffil{einstein}{NHFP Einstein Fellow}
%\DeclareAffil{it}{Dept. Physics and Earth Science, University of Ferrara, via Saragat 1, I-44122, Ferrara, Italy}
%\DeclareAffil{leic}{University of Leicester, Department of Physics \& Astronomy and Leicester Institute of Space \& Earth Observation, University Road, Leicester, LE1 7RH, UK}

\affilauthorlist{K.~D.~Alexander\affils{cfa,nu,einstein}, T.~Laskar\affils{nrao,UCB}, E.~Berger\affils{cfa}, M.~D.~Johnson\affils{cfa}, P.~K.~G.~Williams\affils{cfa},
S.~Dichiara\affils{mex}, W.~Fong\affils{nu,hubble}, A.~Gomboc\affils{slov}, S.~Kobayashi\affils{liver}, R.~Margutti\affils{nu}, C.~G.~Mundell\affils{bath}}

\begin{abstract}
We present Karl G.~Jansky Very Large Array radio observations of the long gamma-ray burst GRB\,161219B ($z=0.147$) spanning $1-37$ GHz.  The data exhibit unusual behavior, including sharp spectral peaks and minutes-timescale large-amplitude variability centered at $20$ GHz and spanning the full frequency range.  We attribute this behavior to scattering of the radio emission by the turbulent ionized Galactic interstellar medium (ISM), including both diffractive and refractive scintillation.  However, the scintillation is much stronger than predicted by a model of the Galactic electron density distribution (NE2001); from the measured variability timescale and decorrelation bandwidth we infer a scattering measure of $SM\approx {(8-70)\times 10^{-4}}$ kpc m$^{-20/3}$ ({up to} ${25}$ times larger than predicted in NE2001) and a scattering screen distance of $d_{\rm scr}\approx {0.2-3}$ kpc. We infer an emission region size of $\theta_s \approx {0.9-4}$ $\mu$as ($\approx {(1-4)}\times 10^{16}$ cm) at $\approx4$ days, and find that prior to 8 days the source size is an order of magnitude smaller than model predictions for a uniformly illuminated disk or limb-brightened ring, indicating a slightly off-axis viewing angle or significant substructure in the emission region. Simultaneous multi-hour broadband radio observations of future GRB afterglows will allow us to characterize the scintillation more completely, and hence to probe the observer viewing angle, the evolution of the jet Lorentz factor, the structure of the afterglow emission regions, and ISM turbulence at high Galactic latitudes.
\end{abstract}


\section{Introduction}

Radio emission from compact sources is distorted as it propagates through the turbulent ionized interstellar medium (ISM) of the Milky Way, producing frequency-dependent flux variations on timescales of minutes to days. This effect, called interstellar scintillation (ISS) \citep{rick90,good97}, has been used to help map the Galactic electron density distribution using pulsars \citep{cl02}. ISS has also been detected in radio observations of sufficiently compact extragalactic sources such as some active galactic nuclei (AGNs), establishing limits on the size of their unresolved compact radio cores to a few tens of microarcseconds \citep{hr87,den02,lov08}, and in transient sources ranging from gamma-ray burst (GRB) afterglows (e.g. \citealt{fra97,fra00,chan08}) to jetted tidal disruption events \citep{bloom11,zaud11} and fast radio bursts \citep{masui15,cor16,katz16}. GRB afterglows are particularly valuable probes of ISS because they can be used to sample high Galactic latitudes, where pulsars are rare and the properties of the turbulent ISM are poorly constrained. Moreover, while AGNs are more common than GRBs across the sky, the generally larger angular sizes of AGNs typically suppress any ISS variability.  GRBs, on the other hand, are initially compact, but also expand with time thereby changing the observed scattering behavior; thus ISS can be used to determine the size evolution of radio-emitting regions in GRBs. With the exception of very long baseline intereferometry (VLBI) observations, which to date have provided strong size constraints for only one event (the nearby GRB 030329, whose radio afterglow remained bright long enough to be resolved starting at $\approx 20$ days post-burst; \citealt{tay04,tay05,pihl07}), ISS is the only method of measuring the sizes of a large sample of GRB afterglows across timescales of days to weeks, providing a direct test of afterglow models. In the case of GRB 970508, the ISS-derived afterglow size provided the first direct confirmation of the now-standard relativistic fireball model for GRBs \citep{fra97,fra00}.

While ISS is expected to be ubiquitous in GRBs, it has only been detected convincingly in a handful of events because previous observations have lacked the bandwidth and cadence needed to characterize the variability in detail.  There are several detections of mild variability with a cadence of days at a single frequency, and only two cases in which variability was tracked for hours, though still at a single frequency \citep{chan08,van14}. {Recently, \cite{greiner18} reported extremely large-amplitude variability in the afterglow of GRB 151027B on timescales of days at two frequencies, possibly requiring a complex distribution of scattering material along the line of sight, but were unable to fully characterize the behavior due to their limited observational coverage.} The large bandwidth and improved sensitivity of NSF's Karl G.~Jansky Very Large Array (VLA) can rectify this situation. Over the past few years, our group has undertaken a systematic study of long GRB afterglows with the VLA, greatly improving the frequency coverage and the temporal sampling at early times. Our observations have revealed a number of unusual features in GRB radio light curves, including reverse shock (RS) emission and novel scattering behavior \citep{las16,lab+18,ale17}. 

Here, we present a study of strong ISS in the radio afterglow of GRB\,161219B. We observe unusually large-amplitude, rapid variability whose strength decreases with time, allowing us to track the size of the afterglow as it expands. Unlike previous ISS detections, which were all below 10 GHz, here the variability peaks at $\approx 20$ GHz, indicating a strongly scattering medium. Our data span $1-37$ GHz, allowing us to place direct constraints on the correlation bandwidth of the observed variability, as well as the variability timescale. Additionally, the brightness of the afterglow allows us finely sample the observations in both time and frequency space, probing variability on timescales of minutes to days in unprecedented detail.  We describe our observations in Section \ref{sec:obs}, define our model for ISS and use it to constrain the properties of the observed scattering medium in Section \ref{sec:iss}, discuss implications for the afterglow size evolution in Section \ref{sec:size}, and conclude in Section \ref{sec:conc}.  We assume standard $\Lambda$CDM cosmology with $H_0 = 68$ km s$^{-1}$ Mpc$^{-1}$, $\Omega_M = 0.31$, and $\Omega_{\Lambda} = 0.69$ throughout.

\section{Radio Observations}
\label{sec:obs}

\begin{figure} 
\vspace{-14pt}
\centerline{\includegraphics[width=0.56\textwidth]{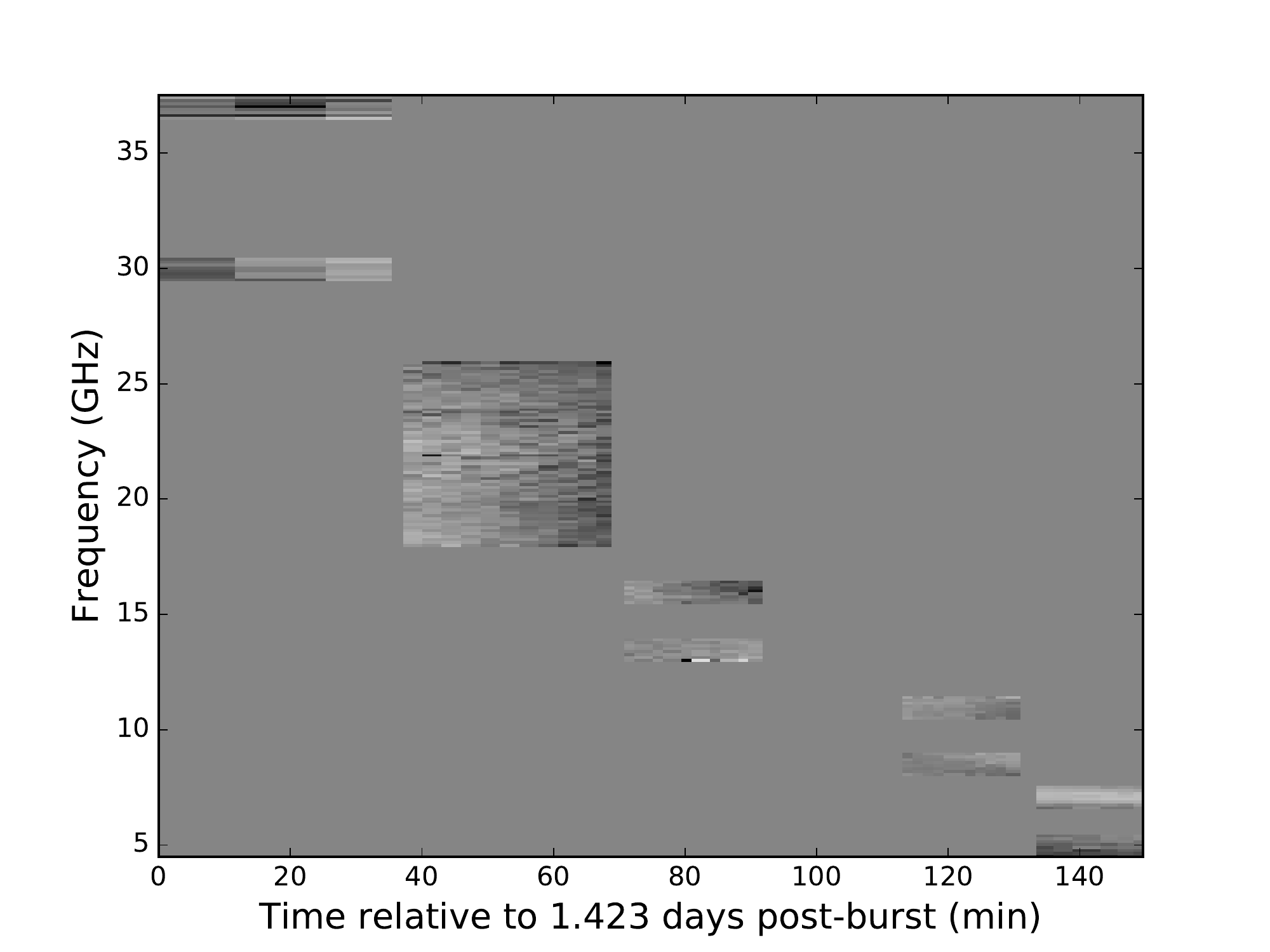}}
\caption{Time$-$frequency ``waterfall" plot showing our frequency coverage as a function of time 1.5 days after the burst. We followed a similar observing strategy in all epochs (see Figures \ref{fig:lc1}$-$\ref{fig:lc5}). The grayscale shows the relative change in flux density in each frequency band (white is $\geq3$ times larger than the mean flux density in each band, black is $\geq3$ times smaller). The changing coherence bandwidth of the short-term variability is clearly visible: the emission at $18-24$ GHz varies coherently and may connect to the trends seen in the 30 and 16 GHz sub-bands, while at lower frequencies the emission in each sub-band varies independently.} 
\label{fig:wf2}
\end{figure}

GRB\,161219B was discovered by the Burst Alert Telescope (BAT; \citealt{bat}) on board the Neil Gehrels {\it Swift} Observatory \citep{swift} on 2016 December 19 at 18:48:39 UT \citep{dai16}. The afterglow and associated Type Ic supernova (SN 2016jca) have been extensively monitored at X-ray through radio wavelengths with a wide range of ground- and space-based facilities (e.g. \citealt{ash17,can17}). Our group obtained the first radio observations of the afterglow at both centimeter (VLA; \citealt{vla}) and millimeter (ALMA; \citealt{alma}) wavelengths. Here, we focus on our early cm-band radio observations at $0.5-16.5$ days. A detailed analysis of the broadband afterglow and a full list of our X-ray, UV, optical, near-IR, millimeter, and centimeter observations are given in a companion publication (\citealt{lab+18}; hereafter LAB18). 

\subsection{Observing Strategy and Data Analysis}
\label{sec:radio}

\begin{center}
\setlength\LTcapwidth{3.5in}
\begin{longtable}{ccccc}
\caption{VLA Radio Observations}
\label{tab:obs} \\
\hline
\hline\noalign{\smallskip}
Epoch & $\Delta t$ & Duration & Receiver & Frequency Range \\
& (days)   & (minutes)   &  &  (GHz)  {\smallskip} \\
\hline\noalign{}
1 & 0.51 & 34 & $K$ & $18-26$ \\
1 & 0.53 & 24 & $Ku$ & $13-14, 15.5-16.5$ \\
1 & 0.55 & 15 & $X$ & $8-9, 10.5-11.5$ \\
1 & 0.56 & 15 & $C$ & $4.5-5.5, 6.6-7.6$ \\
\hline\noalign{}
2 & 1.43 & 41 & $Ka$ & $29.5-30.5, 36.5-37.5$ \\
2 & 1.46 & 31 & $K$ & $18-26$ \\
2 & 1.48 & 20 & $Ku$ & $13-14, 15.5-16.5$ \\
2 & 1.51 & 17 & $X$ & $8-9, 10.5-11.5$ \\
2 & 1.52 & 17 & $C$ & $4.5-5.5, 6.6-7.6$ \\
\hline\noalign{}
3a & 3.56 & 17 & $X$ & $8-9, 10.5-11.5$ \\
3a & 3.57 & 17 & $C$ & $4.5-5.5, 6.6-7.6$ \\
3b & 4.43 & 41 & $Ka$ & $29.5-30.5, 36.5-37.5$ \\
3b & 4.46 & 31 & $K$ & $18-26$ \\
3b & 4.48 & 21 & $Ku$ & $13-14, 15.5-16.5$ \\
\hline\noalign{}
4 & 8.44 & 44 & $Ka$ & $29.5-30.5, 36.5-37.5$ \\
4 & 8.47 & 34 & $K$ & $18-26$ \\
4 & 8.50 & 24 & $Ku$ & $13-14, 15.5-16.5$ \\
4 & 8.51 & 15 & $X$ & $8-9, 10.5-11.5$ \\
4 & 8.52 & 15 & $C$ & $4.5-5.5, 6.6-7.6$ \\
\hline\noalign{}
5 & 16.49 & 34 & $K$ & $18-26$ \\
5 & 16.51 & 24 & $Ku$ & $13-14, 15.5-16.5$ \\
5 & 16.53 & 15 & $X$ & $8-9, 10.5-11.5$ \\
5 & 16.54 & 15 & $C$ & $4.5-5.5, 6.6-7.6$ \\
5 & 16.55 & 15 & $S$ & $2.1-3, 3-3.9$ \\
5 & 16.56 & 23 & $L$ & $1-2$ \\
\hline\noalign{\smallskip}
\caption[]{Summary of the timing, frequency coverage, and VLA receivers used in our GRB 161219B radio observations. For further details see Figures \ref{fig:lc1} $-$ \ref{fig:lc5} and LAB18. All values of $\Delta t$ indicate the mean observation time and are relative to 2016 December 19 18:48:39 UT, the BAT trigger time.}
\end{longtable}
\end{center}
\vspace{-10pt}

We observed the afterglow using the VLA beginning 11.4 hr after the burst under program 15A-235 (PI: Berger). All of the data presented here were obtained in the A configuration. As is standard for VLA observations, we selected one observing band at a time, rotating through receivers sensitive to different frequency ranges from high to low frequency and observed for $15-45$ min in each band (Figure \ref{fig:wf2}). The frequency coverage of each receiver tuning and the timing of each epoch are summarized in Table~\ref{tab:obs}. We used the 3-bit samplers at $K$ band ($18-26$ GHz) to maximize the instantaneous frequency coverage and the 8-bit samplers at other frequencies to maximize sensitivity, with resulting bandwidths of 0.6 GHz at $L$ band ($1-2$ GHz) and 2 GHz at all other frequencies. The usable bandwidth at the lower frequencies ($\lesssim 6$ GHz) was lower than these nominal values due to radio frequency interference (RFI). In all bands except $K$ and $L$, the bandwidth was divided into two sub-bands of 1 GHz each, separated by a gap of up to 1.5 GHz. In the $K$ band, we observed four adjacent sub-bands of 2 GHz each, providing contiguous frequency coverage. In the $L$ band, the two sub-bands were also adjacent, but had gaps in frequency coverage due to RFI.

\begin{figure*} 
\centerline{\includegraphics[width=7.25in]{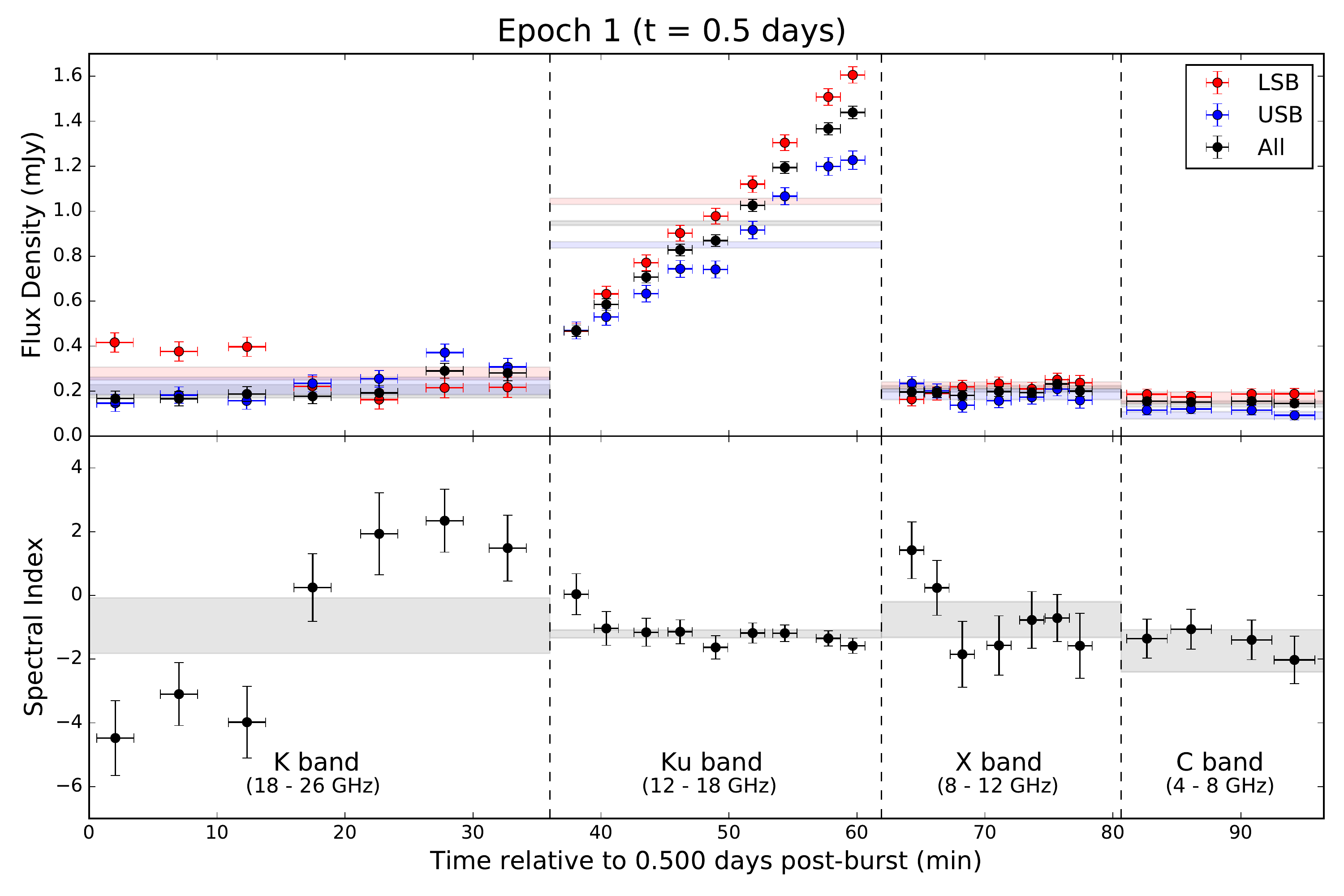}}
\centerline{\includegraphics[height=2.95in]{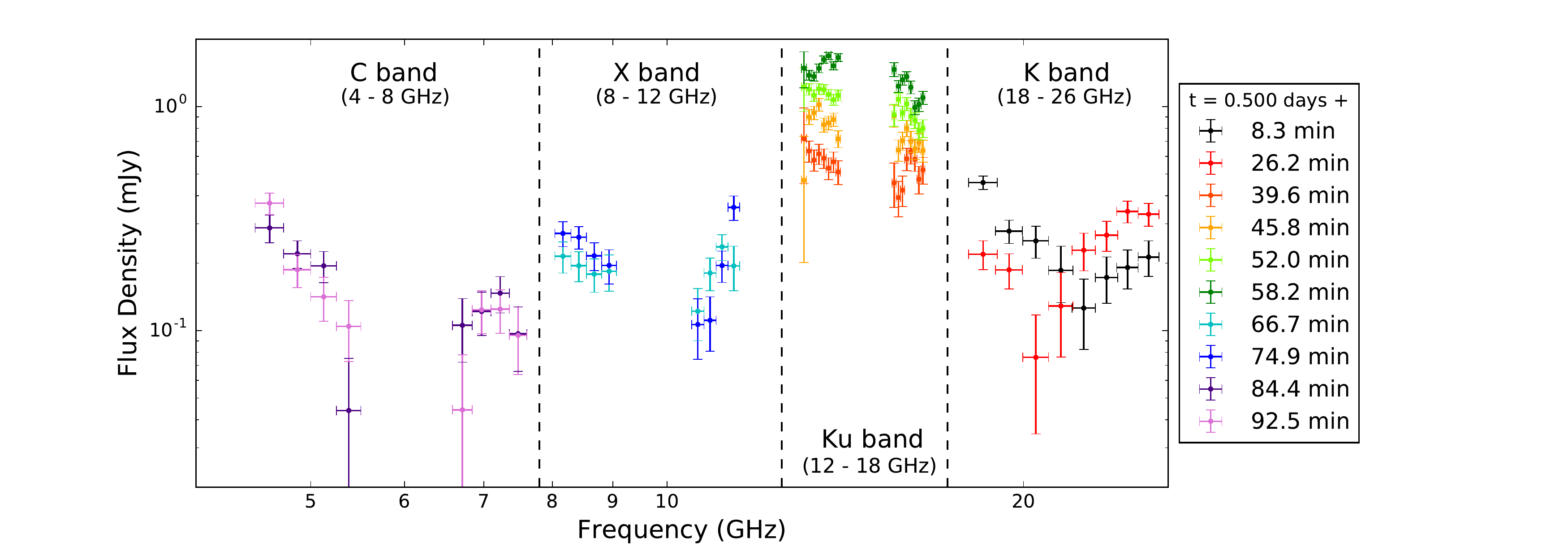}}
\caption{Top: rapid time evolution of the flux density and in-band spectral index during our first epoch of observations at 0.5 days. The upper axis shows the temporal evolution of the lower sideband (LSB; red) and upper sideband (USB; blue) flux densities for each receiver, moving from high frequencies (20 and 24 GHz, $K$ band) to low frequencies (5.0 and 7.1 GHz, $C$ band). The flux density at $Ku$ band increases by a factor of $3-4$ in 24 min. The lower axis shows the spectral index between the USB and LSB for each receiver. The shaded bands show the flux density and spectral index for each receiver obtained from imaging all of the data for each frequency and fitting a point source to the image, as reported in LAB18. Bottom: time-sliced spectral energy distributions (SEDs) at 0.5 days; points of the same color are simultaneous. The SED evolves significantly at the $Ku$ band and marginally at the $K$ band over the duration of our observations. The afterglow is too faint at low frequencies to confirm variability, but the spectral variations at the $C$ and $X$ bands are characteristic of strong ISS. The data used to create this figure are available.}
\label{fig:lc1}
\end{figure*}

\begin{figure*} 
\centerline{\includegraphics[width=7.25in]{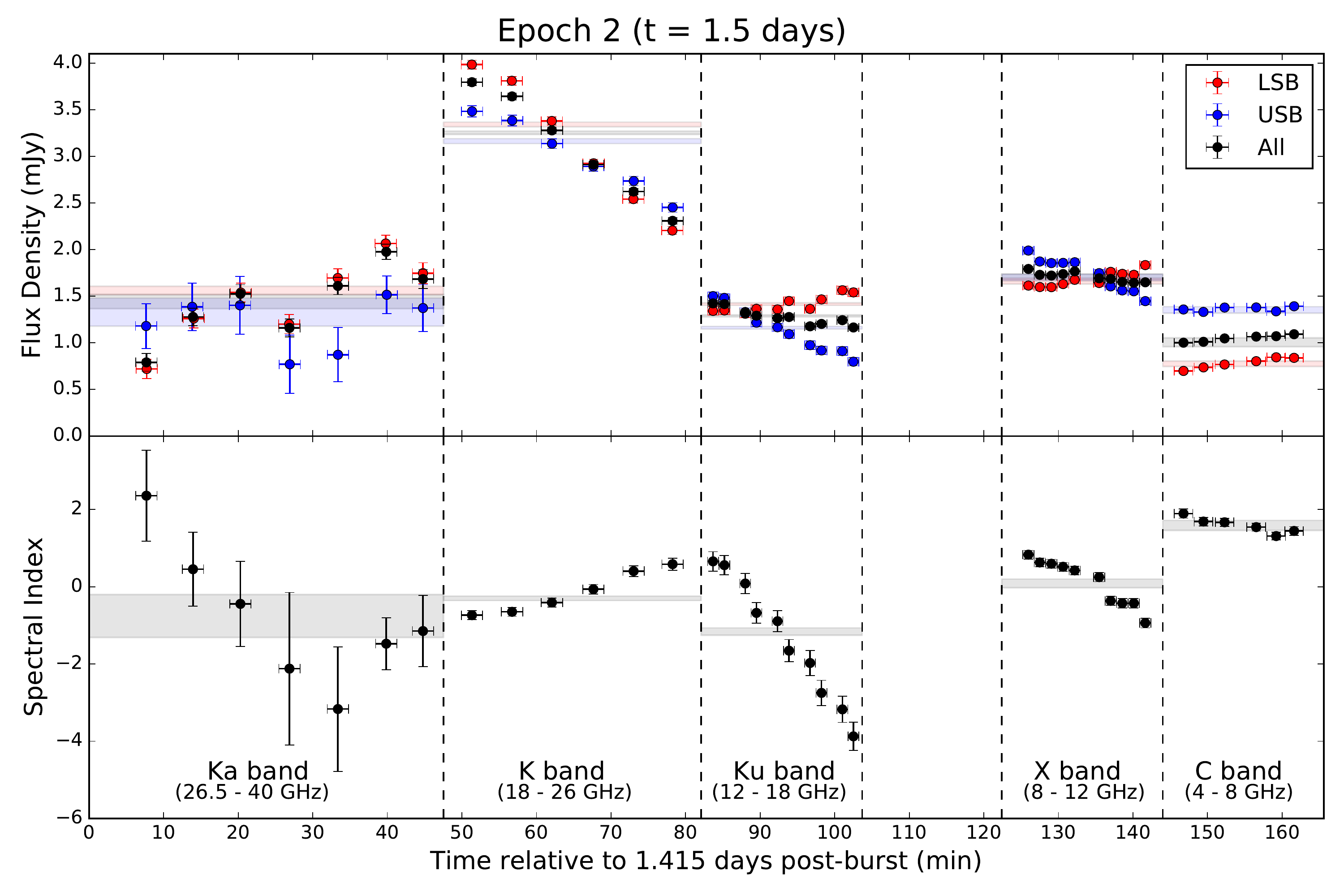}}
\centerline{\includegraphics[height=2.95in]{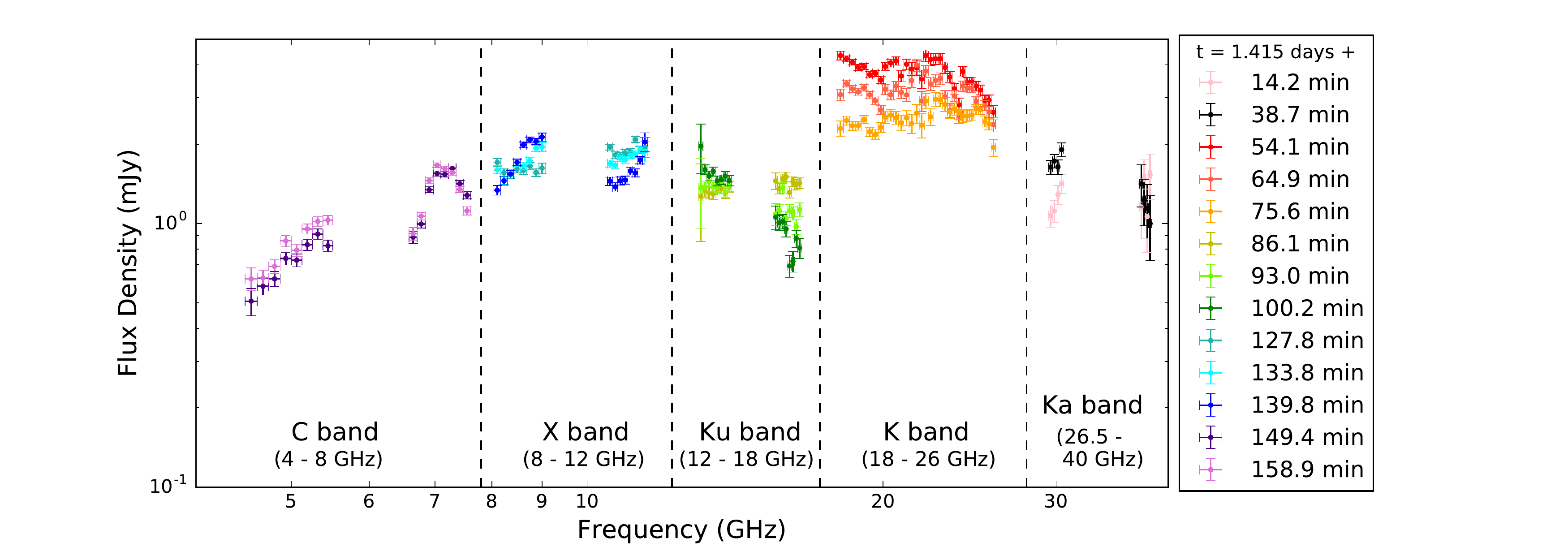}}
\caption{Same as Figure \ref{fig:lc1} for the epoch 2 radio data, showing rapid variability on timescales of tens of minutes 1.5 days after the burst. The largest variations are seen in the $K$ band, suggesting that the transition frequency between strong and weak scattering is $\nu_{\rm ss}\approx22$ GHz. The bottom panel shows that the coherence bandwidth of the variations increases with frequency, as expected for diffractive ISS. Fluctuations are coherent across the full $Ku$ sub-bands at $13-14$ GHz and $15.5-16.5$ GHz, but the coherence bandwidth drops to $\approx500$ MHz by 8.5 GHz. The $C$ band SED does not vary significantly over the duration of the observation, indicating that either diffractive ISS is quenched at frequencies $\simlt 8$ GHz due to a finite source size or that the coherence bandwidth is below the spectral resolution of 128 MHz at these frequencies. The large change in the spectral index at $4.5-5.5$ GHz between 0.5 days and 1.5 days is suggestive of refractive ISS. The data used to create this figure are available.}
\label{fig:lc2}
\end{figure*}

\begin{figure*}
\centerline{\includegraphics[width=7.25in]{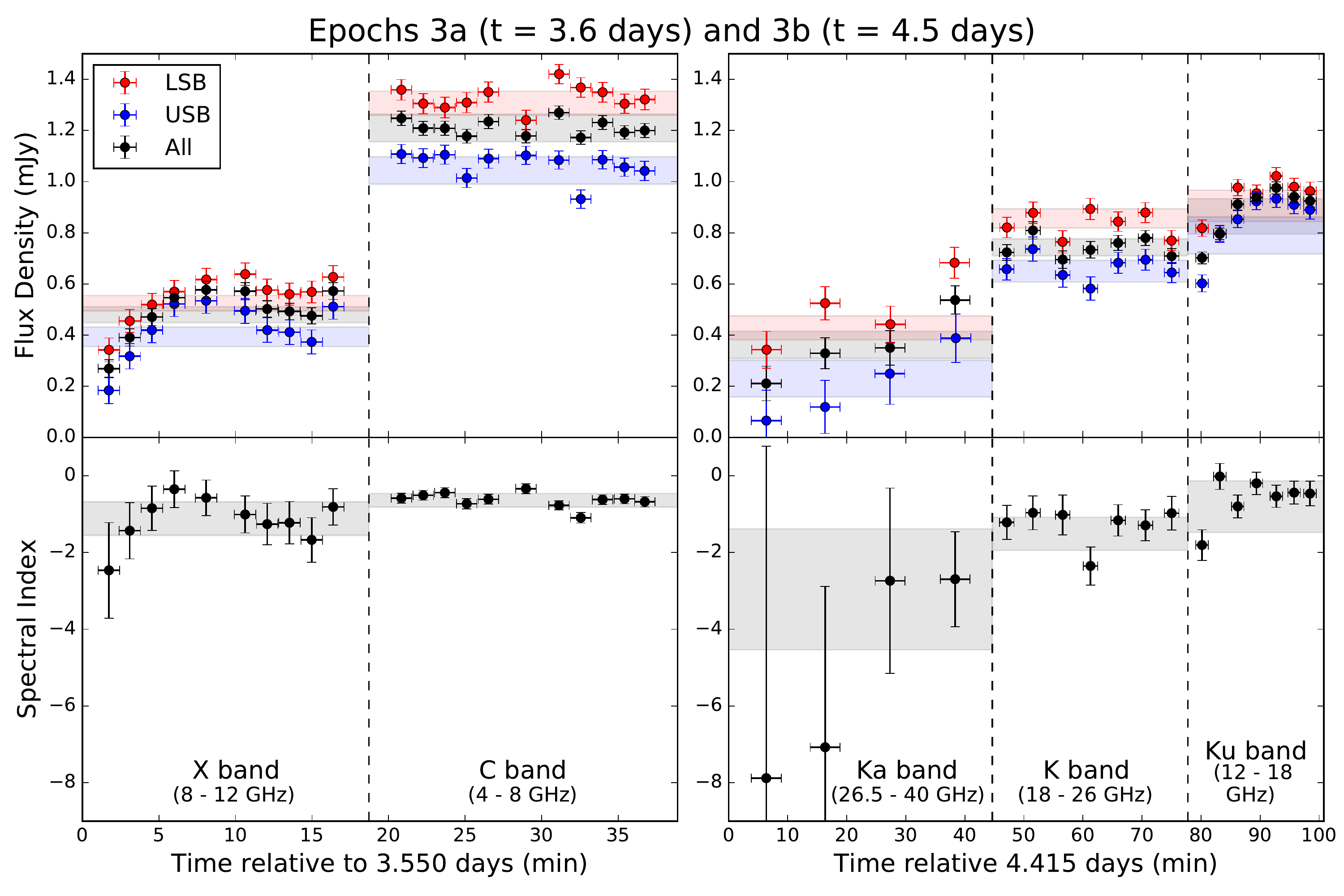}}
\centerline{\includegraphics[height=2.95in]{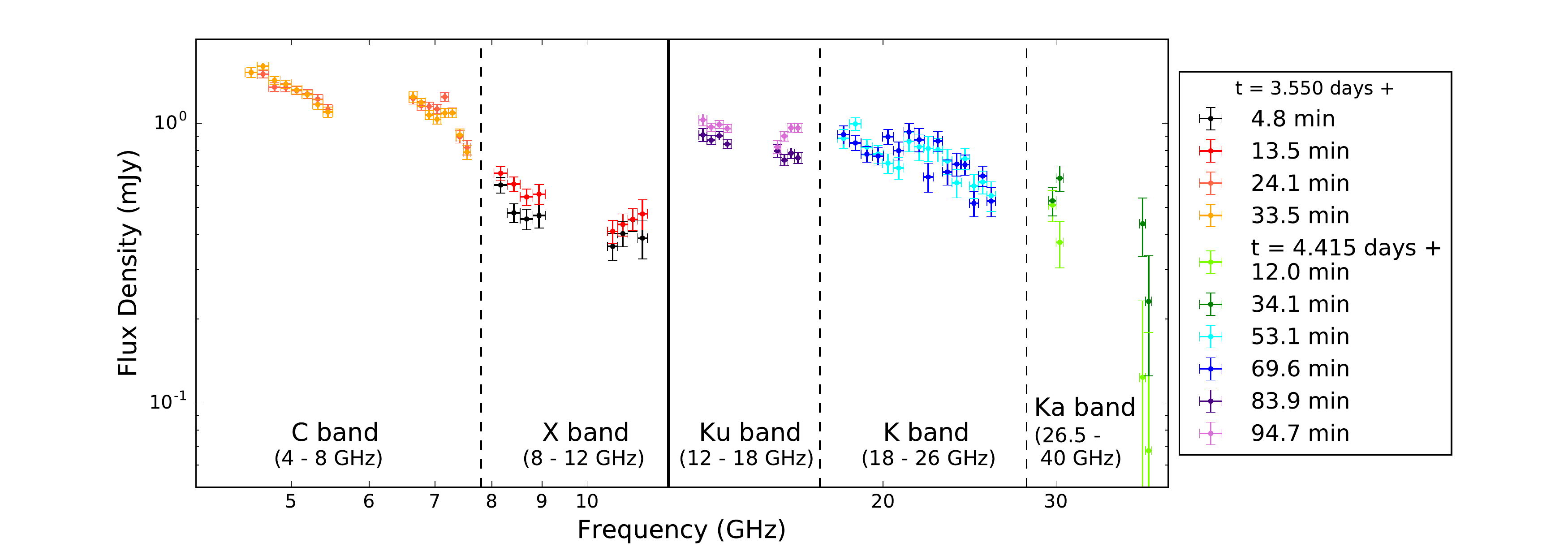}}
\caption{Top: epoch 3a and 3b light curves and spectral index evolution at 3.6 and 4.5 days. Note that the low frequencies ($C$ and $X$ bands; top left) were observed $\sim1$ day earlier than the low frequencies ($Ku$, $K$, and $Ka$ bands; top right). The extreme variability seen in the first two epochs has largely quenched, although there are still hints of variations at the $X$ and $Ku$ bands. Bottom: in the SED plot, the $Ku$ and $X$ bands show weak evidence of variability, but the changes are much less dramatic than in the previous two epochs, indicating that the afterglow is approaching the size limit at which diffractive ISS quenches. The data used to create this figure are available.}
\label{fig:lc3}
\end{figure*}

\begin{figure*} 
\centerline{\includegraphics[width=7.25in]{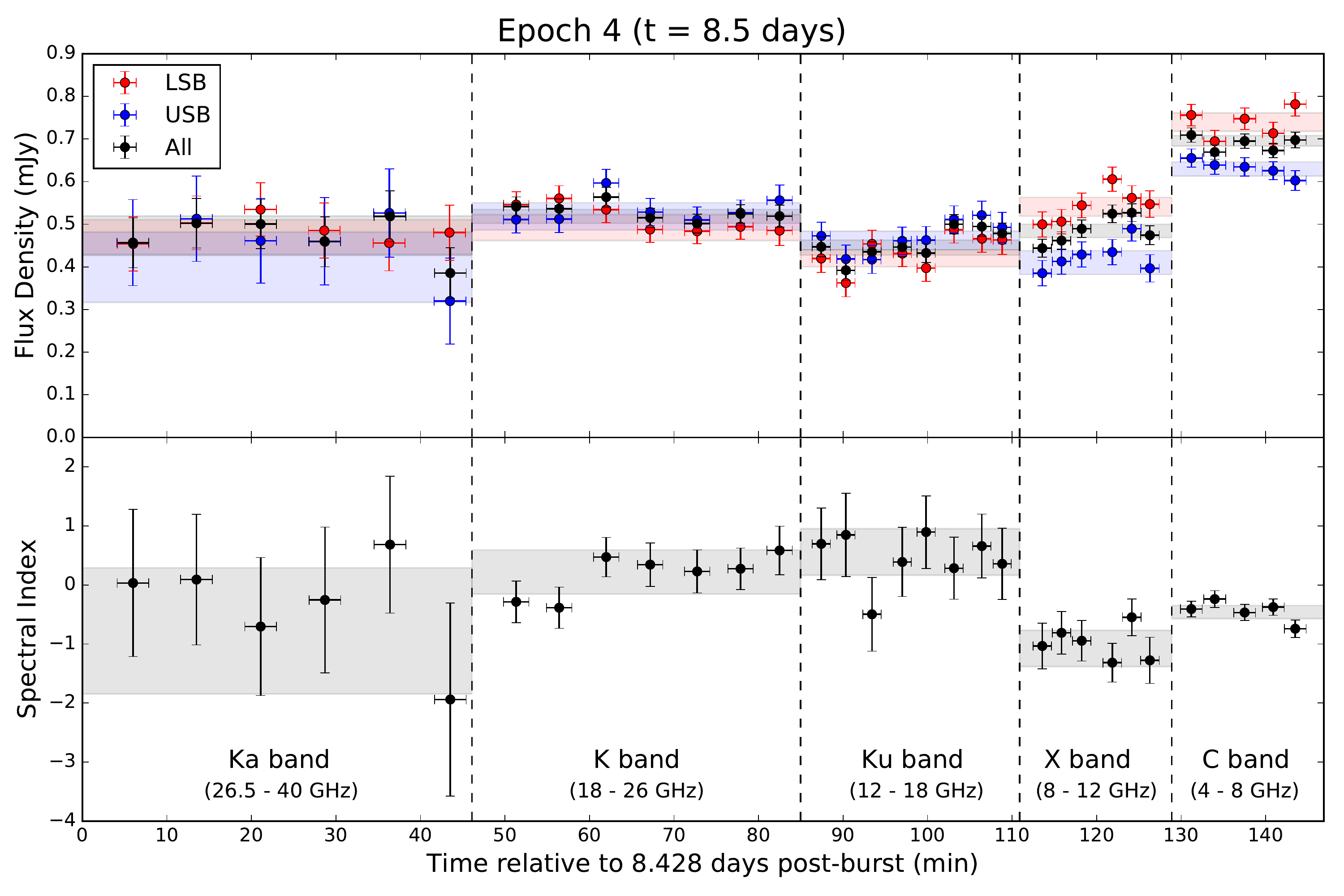}}
\centerline{\includegraphics[height=2.95in]{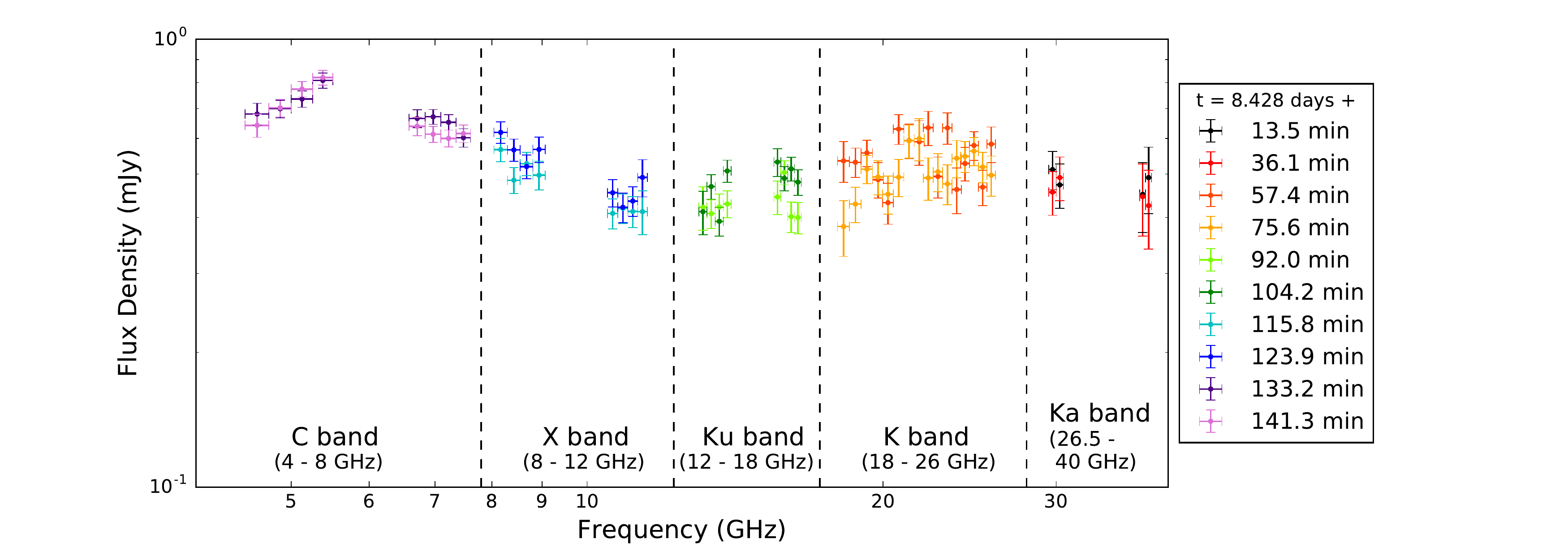}}
\caption{Epoch 4 light curve and SED evolution at 8.5 days. We no longer see any large-amplitude variability within this single observation, but the spectral index in the $C$ band still changes in comparison to the previous epoch at 3.5 days, indicating continuing refractive ISS. The data used to create this figure are available.}
\label{fig:lc4}
\end{figure*}

\begin{figure*} 
\vspace{-0.1in}
\centerline{\includegraphics[width=7.25in]{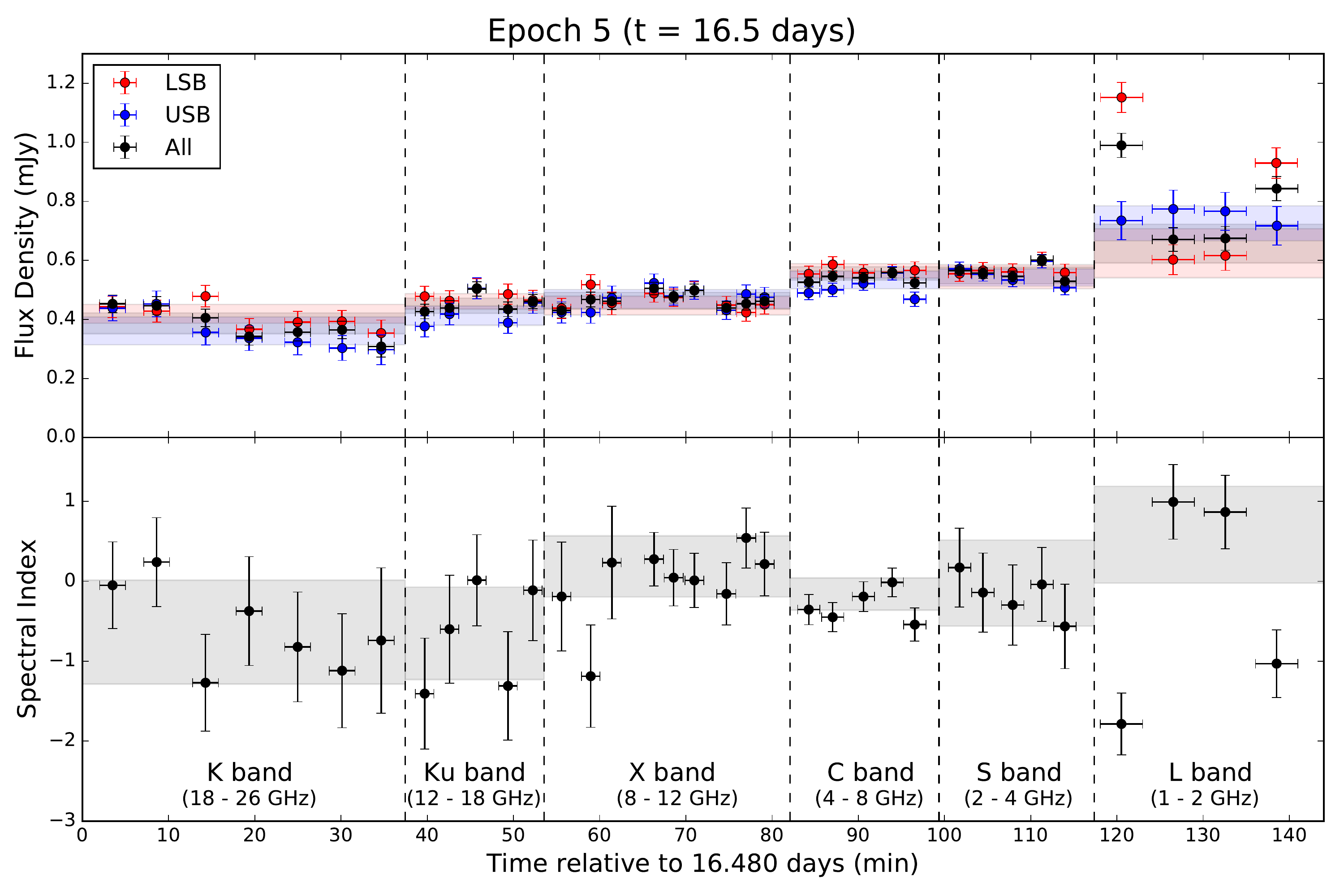}}
\centerline{\includegraphics[height=2.95in]{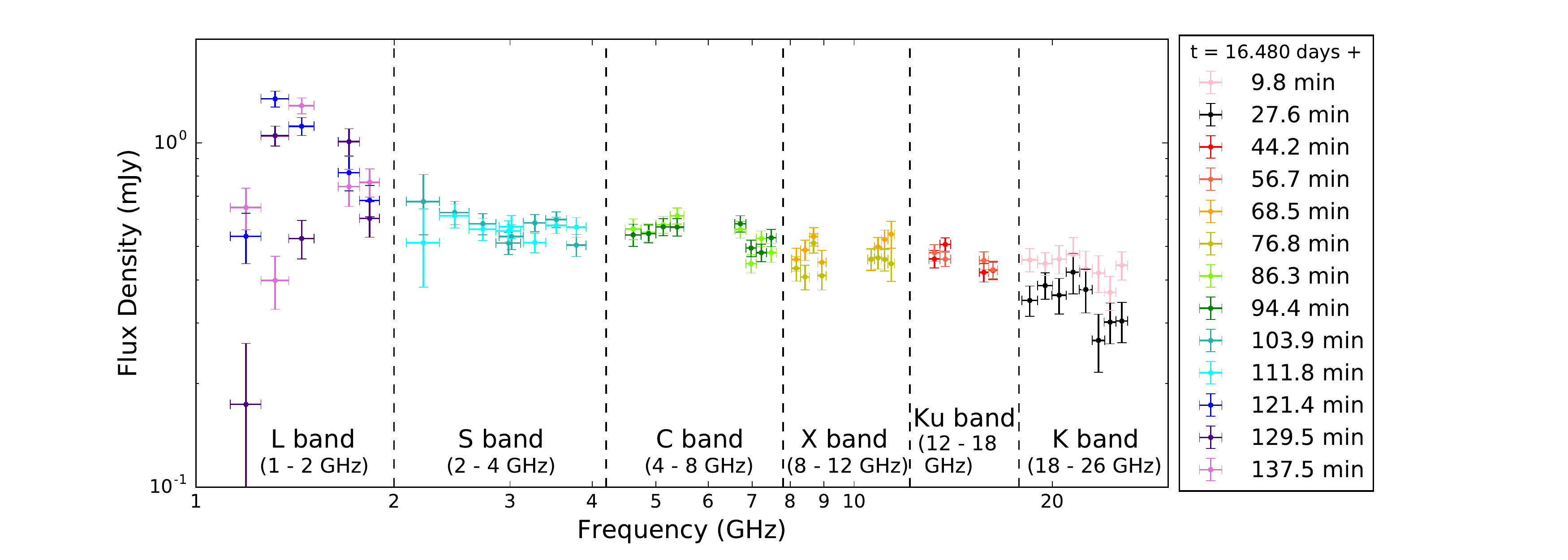}}
\caption{Epoch 5 light curve and SED evolution at 16.5 days. As in epoch 4, we no longer see sharp spectral features or evidence of rapid variability in this epoch, suggesting that refractive ISS has also been suppressed as the afterglow expands. The sharp temporal and spectral discontinuities in the $L$ band LSB ($1-1.5$ GHz) are likely due to data loss from RFI rather than intrinsic variability; RFI minimally affects our measured flux densities at other frequencies. The data used to create this figure are available.}
\label{fig:lc5}
\end{figure*}

We analyzed the data with the Common Astronomy Software Applications (CASA) using 3C48 as a flux calibrator and J$0608-2220$ as a gain calibrator. Initially, we imaged the data using the {\tt CLEAN} algorithm and determined the flux density and associated uncertainties at each band using the \texttt{imtool} program within the \texttt{pwkit} package\footnote{Available at \url{https://github.com/pkgw/pwkit}.} (version 0.8.4.99; \citealt{pwkit}). The flux densities thus obtained are time- and frequency-averaged over the duration and bandwidth of each observation with a particular receiver. They are shown as shaded horizontal bands in Figures~\ref{fig:lc1}--\ref{fig:lc5} (top panels) and are reported in full in LAB18 (their Table 5). 

To probe variability on timescales shorter than the duration of each observation, we used the \texttt{dftphotom} task in \texttt{pwkit} to directly fit the observed visibilities with a point source model centered at the afterglow coordinates using discrete Fourier transforms \citep{pwkit}. The resulting light curves are shown in Figures~\ref{fig:lc1}--\ref{fig:lc5} (top panels). We also tracked the evolution of the spectral index between sub-bands of the same receiver (Figures~\ref{fig:lc1}--\ref{fig:lc5}, middle panels; a positive spectral index indicates increasing flux density with frequency). In addition, we split the data into $128-1024$ MHz frequency segments to track the spectral evolution within each frequency sub-band more precisely (Figures~\ref{fig:lc1}--\ref{fig:lc5}, bottom panels).  We observe large-amplitude flux density and spectral index changes in the first two epochs (0.5 and 1.5 days) at $8-26$ GHz. These effects are strongly diminished in our third epoch (split between 3.6 days and 4.5 days) and disappear before our fourth epoch at 8.5 days.

To demonstrate that residual phase errors in our data do not cause the observed short-term variability, we performed phase-only self-calibration at the $X$ and $Ku$ bands in epoch 1 and in the $Ku$, $K$, and $Ka$ bands in epoch 2. We find that the mean flux density in each band increases by $\approx 10-30$\% after self-calibration, but the intra-epoch variability trends remain unchanged. We show the self-calibrated datasets for these frequencies in Figures~\ref{fig:lc1} and \ref{fig:lc2}.

\subsection{Variability Characteristics}

The rapid temporal variability seen in GRB\,161219B limits our ability to connect features seen in different frequency bands, as the data were not obtained simultaneously (Figure~\ref{fig:wf2}). However, we also see extreme variability within individual frequency bands. For example, the in-band spectral index at 11 GHz at 0.5 days (epoch 1) is an extremely steep $\nu^{12}$ and the flux density at the $Ku$ band in epoch 1 increases by a factor of about $3.5$ in 24 minutes, implying a temporal index of $t^{40}$ (Figure~\ref{fig:lc1}). This corresponds to a brightness temperature $T_{\rm b,obs}\sim10^{18}$ K, which would require superluminal motion along the line of sight with $\Gamma\gtrsim{10^2}$ if the variability is intrinsic to the source \citep{kp69,read94}. {Such a high Lorentz factor is not expected 0.5 days post-burst; \cite{anderson18} found values at least one order of magnitude smaller for a sample of radio-detected GRBs observed at similar epochs.} Additionally, the amplitude of the variability decreases markedly at $3.6-4.5$ days (epoch 3; Figure \ref{fig:lc3}), which is difficult to explain with any mechanism intrinsic to the burst. The high-frequency spectral energy distributions (SEDs) are essentially flat after this time, but we still see unusual behavior at lower frequencies through our fourth epoch at 8.5 days (Figure~\ref{fig:lc4}). Notably, the spectral index within the 5 GHz sub-band ($4.5-5.5$ GHz) changes significantly in each of the first four epochs, from negative to positive to negative to positive. It is only in the final epoch, at 16.5 days, that this trend ceases and all frequencies connect to form a single, smooth SED as expected in the standard afterglow model (Figure~\ref{fig:lc5}; bottom panel).

These sharp spectral features and rapid temporal changes are inconsistent with the intrinsic behavior of GRB afterglows. In the standard picture, the afterglow SED is expected to consist of smoothly connected power-law segments, with the break frequencies and the overall normalization evolving smoothly and moderately in time \citep{gs02}. The intrinsic flux density evolution of the afterglow is slow ($t^{-2}$ at the fastest), so we do not expect to see intrinsic variability on $\lesssim 1$ hr timescales days after the burst. The expected SEDs are broad, with the spectral index varying at most between $2.5$ and $-1.5$ \citep{gs02}. Furthermore, in the simplest model where all of the emission arises from the forward shock (FS), the spectral index in a given band should only evolve from positive to negative, not undergo repeated sign flips as we observe at 5 GHz at $0.5-8.5$ days. These spectral index changes cannot be explained even in the context of a more complex FS plus RS model because the implied RS evolution is too fast; LAB18 predict that the RS component should entirely dominate the emission at 5 GHz until 8.5 days. Below, we show that the extreme features at early times can be explained as diffractive ISS (DISS), while the broadband variability at lower frequencies and later times is due to refractive ISS (RISS).

\section{Analytic Scattering Model}
\label{sec:iss} 

We first provide a basic overview of analytic scattering theory as it applies to GRB\,161219B. (For a more complete treatment of this topic, see \citealt{rick90}.) The characteristic angle by which incoming light rays are scattered while traversing the ISM depends on frequency and on the amplitude of the electron density inhomogeneities encountered along the line of sight, which is quantified by the scattering measure, $SM$. If this scattering angle is small, then only a single image of the source is produced and the resulting flux variations are small (weak scattering). Conversely, if the scattering angle is large, then multiple images of the source are formed and the flux can vary significantly (strong scattering). In both strong and weak scattering, the received flux varies across the observer plane due to the focusing and defocusing of individual images by inhomogeneities in the scattering medium. In the strong scattering regime, this is called RISS (Section \ref{sec:riss}) and is one of two important scattering processes. In the other, DISS (Section \ref{sec:diss}), light rays emitted from the same point that take different paths to reach the observer interfere to produce a speckle pattern in the observer plane. This speckle pattern is smeared for incoherent radio sources with an angular size larger than the typical speckle size, strongly suppressing the observed variability, so DISS can be used to set an upper limit on the source size if observed (Section \ref{sec:size}). RISS is also suppressed for insufficiently compact sources, but the resulting source size limit is not as stringent. DISS produces the largest amplitude variations (of order unity), but is strongly frequency dependent and may appear suppressed at low frequencies due to frequency-averaging of the data. RISS produces smaller modulations but is a broadband effect.

In the following discussion we ignore scattering within the GRB host galaxy and in the intergalactic medium, as these are expected to be negligible compared to scattering by the Milky Way ISM \citep{good97}. Scattering by the ISM of an intervening galaxy along the line of sight to the GRB might be significant, but no such system has been observed for GRB\,161219B and optical spectra of the afterglow show absorption lines only at the GRB redshift of $z=0.1475$ \citep{deu16,tan16,ash17,can17}. To simplify the discussion, we make the standard assumption that all of the scattering occurs within a thin screen located at a distance $d_{\rm scr}$ from the observer. In this case, strong scattering occurs at all frequencies $\nu < \nu_{\rm ss}$, where \citep{good97}
\begin{equation}\label{eq:ss}
\nu_{\rm ss} \equiv 10.4(SM_{-3.5})^{6/17}d_{\rm scr, kpc}^{5/17}\,\,{\rm GHz},
\end{equation}
where $SM_{-3.5}\equiv SM/(10^{-3.5}\,\,{\rm kpc}\,\, {\rm m}^{-20/3})$ and $d_{\rm scr}$ is in units of kpc. We focus our discussion below on strong scattering, as the large measured flux density variations indicate that is the relevant regime for our observations of GRB\,161219B. 

We use the NE2001 model of the Galactic distribution of free electrons \citep{cl02} as a starting point to estimate the effects of ISS on our observations. As this model is constrained largely by pulsar observations, it is less reliable away from the Galactic plane. For the line of sight to GRB\,161219B (Galactic coordinates $\ell= 233.14592^{\circ}$, $b=-21.04465^{\circ}$), NE2001 predicts $SM_{-3.5}\approx 0.8$ and $\nu_{\rm ss}=12.1$ GHz, leading to $d_{\rm scr}\approx 2.1$ kpc. This is clearly inconsistent with our observations, as the large flux variations in epoch 2 imply that the strong scattering regime extends up to $\nu_{\rm ss,obs} \approx 20-25$ GHz (Figure~\ref{fig:lc2}). If we assume that the NE2001 model correctly determines the $SM$, then Equation~\ref{eq:ss} requires $d_{\rm scr}\approx 12-25$ kpc. This is physically implausible because it would place the scattering screen in the Galactic halo, rather than the disk where most of the scattering material is located. We therefore conclude that the NE2001 model is unreliable for the line of sight to GRB\,161219B and instead estimate $SM$ and $d_{\rm scr}$ directly from our observations.

% check for pulsars

%At frequencies below $\nu_{\rm ss}$, we observe both diffractive and refractive scintillation effects. Diffractive scintillation refers to the interference pattern generated among small patches of coherent phase variations. Refractive scintillation refers to the focusing and defocusing of individual images by larger inhomogeneities in the scattering medium, producing flux increases and decreases. If the observer is moving with respect to the scattering screen, then the flux received at the telescope will undergo random variations as the line of sight passes through this pattern of bright and dark spots. 

\subsection{Diffractive ISS}
\label{sec:diss}

In the first two radio epochs the measured flux density changes by up to a factor of 4 within the time spent observing at a single frequency ($15-45$ minutes). The rapid timescale of these variations along with their large amplitude implies that they are caused by DISS. The timescale for DISS variations is determined by the observer's transverse motion with respect to the scattering screen ($v_{\perp}$) and is defined to be the time it takes for the line of sight to cross a typical diffraction speckle \citep{good97}:
\begin{equation}\label{eq:tdiff}
t_{\rm diff} = 3.1\nu_{10}^{6/5}(SM_{-3.5})^{-3/5}\left(\frac{v_{\perp}}{30\,\,\text{km s}^{-1}}\right)^{-1}\,\, {\rm hr}.
\end{equation}
For our analysis, we assume that $v_{\perp}$ is dominated by the Earth's motion relative to the local standard of rest and is therefore a known quantity. For the line of sight to GRB\,161219B at the time of our observations this motion is $v_{\perp}=31$ km s$^{-1}$.

DISS variations are correlated over a bandwidth that scales with frequency as \citep{good97}:
\begin{equation}\label{eq:dv}
\Delta\nu \approx 7.6 \nu_{10}^{22/5}(SM_{-3.5})^{-6/5} d_{\text{scr,kpc}}^{-1} \text{ GHz}.
\end{equation}
Near $\nu_{\rm ss}$, the correlation bandwidth is comparable to the observing frequency, $\Delta\nu/\nu\approx 1$, while at lower frequencies $\Delta\nu$ rapidly declines below the frequency resolution of our observations and the flux variations from DISS are therefore strongly suppressed. We can see this effect most clearly in epoch 2 (Figure~\ref{fig:lc2}). The variability appears minimal in the $X$ and $Ku$ bands when all of the data in each band are imaged together, but sharp spectral features are revealed when the data are binned more narrowly in frequency. Furthermore, we see no signs of spectral variability at lower frequencies within this epoch, because at frequencies $\nu \lesssim 8$ GHz, $\Delta\nu$ drops below 128 MHz (the narrowest frequency binning possible with our data). 

From Equation~\ref{eq:tdiff}, $t_{\rm diff}$ is directly tied to the $SM$. In long observations, $t_{\rm diff}$ can be determined directly from the observations by constructing intensity structure functions (e.g., \citealt{chan08}). Unfortunately, we do not observe with any single receiver long enough to measure a complete variability cycle; for each frequency we see only monotonic increases or decreases in flux density in each epoch, not random oscillations about a mean value. Therefore, we can only place lower limits on $t_{\rm diff}$ as a function of frequency, giving an upper limit on the scattering measure. The tightest constraint comes from our $X$ band observations in epoch 2, where we have $t_{\rm diff}\gtrsim 17$ min at $8-9$ GHz, or $SM_{-3.5} \lesssim 40$. From Equation~\ref{eq:ss}, this gives us a screen distance $d_{\rm scr}\gtrsim 0.1$ kpc. Here, we are limited by both the uncertainty on $t_{\rm diff}$ and by that on our measurement of the transition frequency, $\nu_{\rm ss}\approx 20-25$ GHz. This is the closest $d_{\rm scr}$ allowed by the data, indicating that the dominant scattering material is at most $\approx 20$ times closer than predicted by NE2001. %We refer to $\nu_{\rm ss}\approx 22$ GHz, $SM_{-3.5} \approx 40$, and $d_{\rm scr} \approx 0.2 $ kpc as the ``Close Screen Model" for the rest of the paper.

%The best way to improve these constraints is to collect longer observations with a larger simultaneou$S$ bandwidth for future events. 
We next explore whether it is possible to improve these constraints by connecting the variability in adjacent frequency bands. {In epoch 2, the flux decline seen in the $K$ band at $18-26$ GHz appears to continue in the upper $Ku$ sideband observations taken at $15.5-16.5$ GHz immediately afterwards (Figure \ref{fig:lc2}). Indeed, because $\Delta\nu/\nu\approx 1$ near $\nu_{\rm ss}$ and $\nu_{\rm ss}\approx 20-25$ GHz, we expect to see coherent variations over this frequency range. We therefore can constrain $t_{\rm diff} \gtrsim 70$ min at $\nu\approx21$ GHz. From equations \ref{eq:ss} and \ref{eq:tdiff}, this gives $SM_{-3.5}\lesssim 20$ and $d_{\rm scr}\gtrsim 0.2$ kpc, improving our constraints on these quantities by a factor of 2. For the rest of this paper, we assume that both $SM$ and $d_{\rm scr}$ are constant in time.}

Although we cannot place an upper limit on ${t_{\rm diff}}$ from our observations directly, we can use our knowledge of the likely distribution of Galactic scattering material to put a soft upper limit on ${d_{\rm scr}}$, which then allows us to compute an upper limit on ${t_{\rm diff}}$ and a lower limit on ${SM}$. The scale height of diffuse, ionized gas in the Milky Way is $\sim1$ kpc in the solar neighborhood (e.g. \citealt{cl02}). For the Galactic latitude of GRB 161219B ($-21^{\circ}$), the assumption that the scattering material is located within one scale height of the Galactic plane gives ${d_{\rm scr} \simlt 3}$ kpc. From equations \ref{eq:ss} and \ref{eq:tdiff}, we then obtain ${SM_{-3.5} \simgt 3}$ and ${t_{\rm diff} \simlt 4}$ hr at 21 GHz for ${\nu_{ss}=20}$ GHz (${SM_{-3.5} \simgt 5}$ and ${t_{\rm diff} \simlt 3}$ hr for ${\nu_{ss}=25}$ GHz).

%Because $\Delta\nu/\nu\approx 1$ near $\nu_{\rm ss}$, for $\nu_{\rm ss}\approx 20-25$ GHz we expect the upper half of $Ku$ band, all of $K$ band, and all of $Ka$ band to vary coherently. In Figure~\ref{fig:lc2} we appear to see a full variability cycle, from minimum to maximum back to minimum, over the 104 minute combined duration of these observations. Taking $t_{\rm diff}\approx 104$ min and $\nu_{\rm ss}\approx 22$ GHz, we use Equations~\ref{eq:ss} and \ref{eq:tdiff} to find $SM_{-3.5}\approx 12$ and $d_{\rm scr}\approx 0.6$ kpc. This model is consistent with all observed frequency dependencies of $\Delta\nu$ and $t_{\rm diff}$ in epochs 1 and 2. For example, it predicts that diffractive fluctuations at 11 GHz (the upper sub-band in $X$ band) should have $\Delta\nu_{\rm dc}\approx 1$ GHz and $t_{\rm diff}\approx 45$ min. This means that we should see a partial variability cycle in Figure~\ref{fig:lc2}, and indeed we see the flux across most of this 1 GHz  sub-band decrease by $\approx 40\%$ in 17 min. We consider $\nu_{\rm ss}\approx 22$ GHz, $SM_{-3.5}\approx 12$, and $d_{\rm scr}\approx 0.6$ kpc to be the most likely scattering model for our observations and refer to it hereafter as the ``Fiducial Model.''

%In addition to the above ``best-fit" model, it is also instructive to consider what range of $SM$ and $d_{\rm scr}$ are allowed.

% give estimate of t_quench here, defer remaining discussion to sect. 4

\subsection{Refractive ISS}\label{sec:riss}

The rapid DISS variability described in the previous section is strongly suppressed by the third epoch at 3.5 and 4.5 days, and by 8.5 days we no longer see variability within individual observations. However, even after DISS quenches at $\approx 4$ days, we continue to observe slower variability in the radio light curves. The dominant effect is a slow fading at all frequencies, which is intrinsic to the GRB afterglow evolution (LAB18), but the spectral index within bands varies non-monotonically, which is a sign of continuing ISS. This behavior is most obvious within the 1 GHz sub-band centered at 5 GHz (Figures~\ref{fig:lc1}--\ref{fig:lc4}). These variations are too broadband to be produced by DISS (Equation~\ref{eq:dv}), but are plausible for RISS.  

At early times, when DISS still dominates the variability at $\gtrsim 8$ GHz, the afterglow can be approximated as a point source for the purposes of characterizing RISS. The characteristic RISS timescale for a point source in the strong scattering regime is \citep{good97}:
\begin{equation}
\begin{split}
t_{\rm ref} &= 4.1\, \nu_{10}^{-11/5} (SM_{-3.5})^{3/5} {d_{\rm scr, kpc}} \\ 
&\: \times \left(\frac{v_{\perp}}{30\text{ km s}^{-1}}\right)^{-1} \text{ hr}
\end{split}
\end{equation}
and the root-mean-square amplitude of the fluctuations is characterized by the modulation index \citep{good97}:
\begin{equation}
m_{\rm ref} = 0.477 \nu_{10}^{17/30} (SM_{-3.5})^{-1/5} d_{\rm scr, kpc}^{-1/6}.
\end{equation}

Our inferred values of $SM$, $d_{\rm scr}$ and $v_{\perp}$ (Section \ref{sec:diss}) imply that at 5 GHz $t_{\rm ref}\approx {20-140}$ hr and $m_{\rm ref}\approx 0.2$, consistent  with the lack of variability seen at this frequency on timescales of tens of minutes. {The lower end of this range, corresponding to the highest allowed $SM$ values and the smallest $d_{\rm scr}$ values, is most consistent with the spectral inversion at 5 GHz that occurs between epochs 1 and 2 (taken 22 hr  apart); this likely means that $d_{\rm scr} \ll 3$ kpc (i.e. the scattering material is well within the Galactic disk), $SM_{-3.5}$ is close to the maximum allowed value of 20, and $t_{\rm diff}$ is much closer to 70 min than 4 hr. However, we retain the full parameter ranges throughout this paper to be conservative.} We continue to observe changes in the spectral index at 5 GHz through 8.5 days (Figure~\ref{fig:lc4}), but at 16.5 days the afterglow no longer shows substantial spectral or temporal variability (epoch 5; Figure~\ref{fig:lc5}), suggesting that the effects of RISS have decreased compared to our earlier epochs. We consider the implications of this in the next section. 

%The modulation index $m_R$ peaks in the strong scattering regime at a frequency given by \citep{good97}:
%
%\begin{eqnarray}
%\nu_{\rm peak} &=& 3.7 \left(\frac{\theta_s}{10\mu{\rm as}}\right)^{-5/11}(SM_{-3.5})^{3/11} \text{ GHz} \\
%m_{R,{\rm peak}} &=& 0.35\left(\frac{\theta_s}{10\mu{\rm as}}\right)^{-17/66}(SM_{-3.5})^{-1/22} d_{\rm scr,kpc}^{-1/6}
%\end{eqnarray}
%where $\theta_s$ is the intrinsic source size. In our data, the modulation index peaks at $\sim20$ GHz. Our individual epochs are not long enough to capture a full cycle of the variability at these frequencies, so we set lower limits of $m_R \gtrsim 0.41$ and $t_{\rm ref} \gtrsim 0.5$ hr on the on the modulation index and variability timescale. 

\section{ISS Constraints on Source Size and Outflow Geometry}
\label{sec:size}

The observed variability allows us to constrain the physical size of the afterglow at multiple epochs, enabling a direct comparison to the afterglow model presented in LAB18. DISS can only produce observable flux variations if the source angular size, $\theta_s$, satisfies \citep{good97}:
\begin{equation}
\label{eq:thetas}
%{\theta_{s} \lesssim 2.25\nu_{10}^{6/5}(SM_{-3.5})^{3/5} \left(\frac{\nu_{\bf ss}}{10.4 { GHz}}\right)^{-17/5}\,\, \mu{\bf as}.} 
\theta_{s} \lesssim 2.25\nu_{10}^{6/5}(SM_{-3.5})^{-3/5} d_{\rm scr, kpc}^{-1}\,\, \mu{\rm as}.
\end{equation}
This limit becomes increasingly restrictive at low frequencies, so if we observe an abrupt cutoff in DISS then we can use it measure the source size (or set an upper limit, if DISS instead cuts off due to $\Delta\nu$ declining below our frequency resolution; Equation~\ref{eq:dv}). In epoch 2 we observe clear variability down to $\approx 8$ GHz. We can therefore set a limit of $\theta_s\lesssim {1}$ $\mu$as at 1.5 days {for ${SM_{-3.5}=20}$, ${d_{\rm scr}=0.2}$ kpc. (Smaller values of $SM$ and larger $d_{\rm scr}$ require a smaller $\theta_s$ for a given cutoff frequency; ${SM_{-3.5}=3}$ and ${d_{\rm scr}=3}$ kpc give $\theta_s\lesssim {0.3}$ $\mu$as.)}

From Equation~\ref{eq:ss}, the maximum frequency at which we observe DISS is $\nu_{\rm ss}$.  Combining this with Equation~\ref{eq:thetas}, we see that DISS is quenched at all frequencies if the source is larger than a critical angular size $\theta_s > \theta_{\rm crit}$ \citep{good97}:
\begin{equation}
\theta_{\rm crit} = 2.35(SM_{-3.5})^{-3/17}d_{\rm scr, kpc}^{-11/17}\,\, \mu{\rm as}.
\end{equation}
For the {constraints given in Section \ref{sec:diss}}, we find $\theta_{\rm crit} \approx {0.9-4}$ $\mu$as. GRB afterglows expand with time, so we expect to see DISS quench at all frequencies when the angular size of the emitting region exceeds $\theta_{\rm crit}$. This naturally explains the transition from the large intra-epoch flux variations and sharp spectral features seen in epochs 1 and 2 to the slower, gentler variability seen subsequently, suggesting that DISS quenches at $t_{\rm crit}\approx 4$ days, and hence $\theta_s\approx {0.9-4}$ $\mu$as at 4 days.

RISS provides no independent information on the source size in the DISS regime, but after $t_{\rm crit}$ we can no longer treat the afterglow as a point source and the modulation index decreases in direct proportion to the source size, $m_{\rm ref}\propto \theta_s^{-7/6}$ \citep{good97}.  In this regime, $m_{\rm ref}$ peaks at a frequency $\nu_{\rm p, ref}$ given by \citep{good97}:
\begin{equation}
\label{eq:vp}
\nu_{\rm p, ref} = 3.7 \left(\frac{\theta_s}{10 \text{ $\mu$as}}\right)^{-5/11} (SM_{-3.5})^{3/11} \text{ GHz.}
\end{equation}
In principle, we can use $m_{\rm ref}$ to measure the source size in all epochs after 4 days, but in practice at late times GRB\,161219B's afterglow is too faint and our cadence is too sparse to place useful independent constraints. However, we can make use of Equation~\ref{eq:vp} in epoch 4, where the only obvious evidence of RISS is at low frequencies, suggesting $\nu_{\rm p, ref}\approx 4-8$ GHz. This suggests that the afterglow size is $\theta_s\approx {3-50}$ $\mu$as at 8.5 days. 

The uncertainty on the first two size measurements is determined by how well we can constrain $SM$ and $d_{\rm scr}$, while the third measurement {additionally} depends on $\nu_{\rm p,ref}$. We assume that $SM$ and $d_{\rm scr}$ are constant in time and compute them from our observables $\nu_{\rm ss}$ and $t_{\rm diff}$ using equations \ref{eq:ss} and \ref{eq:tdiff}. 
The uncertainty on the first two size measurements is dominated by our limited ability to constrain $t_{\rm diff}$, although the uncertainty in $\nu_{\rm ss}$ also contributes. The much larger uncertainty on the final measurement at 8.5 days is due primarily to the strong dependence of $\theta_s$ on $\nu_{\rm p, ref}$, which is only constrained to a factor of $\approx 2$ by our observations. In all epochs, the largest allowed ${\theta_{s}}$ corresponds to the largest allowed value of $SM$, and thus ultimately to the smallest $t_{\rm diff}$ allowed by the data. Therefore, the maximum ${\theta_{s}}$ in each epoch is directly determined by our observations and does not depend on any assumptions made about Galactic structure (our assumed $d_{\rm scr}$ upper limit in Section \ref{sec:diss} provides a lower limit on $SM$ and lower limits on ${\theta_{s}}$).

\begin{figure}
\centerline{\includegraphics[width=0.55\textwidth]{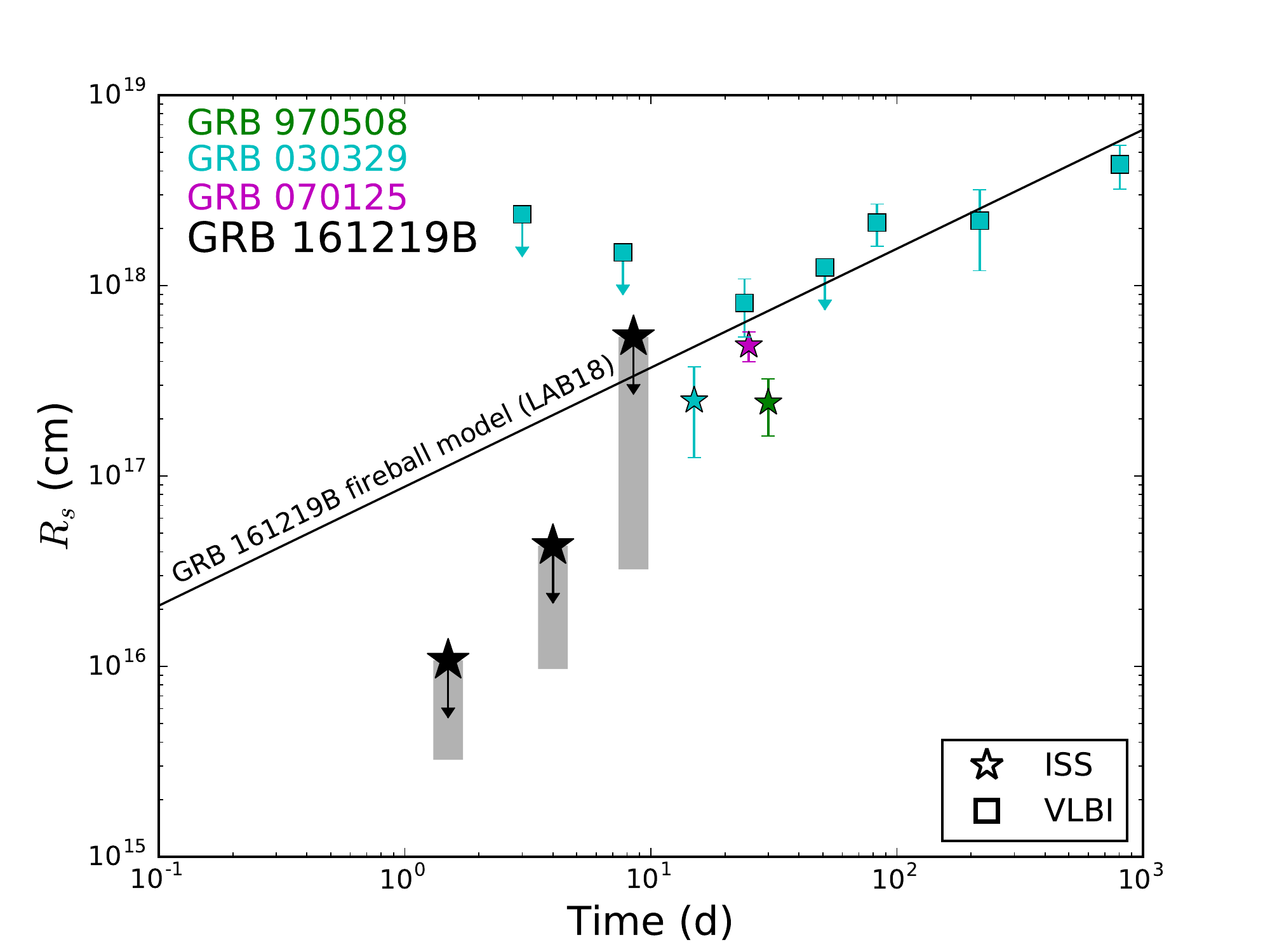}}
\caption{Constraints on GRB afterglow sizes from the literature (colored points; \citealt{fra00,berg03,tay04,tay05,pihl07,chan08}) in comparison to those derived in this work (gray rectangles).  We note that $R_s$ denotes the transverse size of the afterglow image on the sky, not the radial distance from the point of explosion. Squares indicate size measurements and upper limits from VLBI observations, while stars are estimates from ISS. Our ISS results for GRB 161219B are shown together with the predicted size evolution for the fireball model presented in LAB18, which assumes that the afterglow image is a uniformly illuminated disk (black line). Solid black stars indicate ISS size {upper limits derived from our direct observation that $t_{\rm diff}>70$ min at 21 GHz}, while the shaded gray regions show the full range of sizes allowed for $\nu_{\rm ss}\approx20-25$ GHz, ${d_{scr}<3}$ {kpc}, and our constraints on $\nu_{\rm p,ref}$. Even if the scattering properties are pushed to the limit of what is allowed by the data, the discrepancy between the ISS and LAB18 size estimates at early times cannot be reconciled. This may imply substructure in the outflow or a mildly off-axis viewing geometry.}
\label{fig:size}
\end{figure}

Figure \ref{fig:size} shows all three size measurements (shaded gray regions) in comparison to the afterglow model presented in LAB18 (black line) and to size estimates of other GRBs in the literature (colored points). {The black stars indicate the maximum afterglow size allowed by our observations; our early RISS observations at 5 GHz suggest that the true size is closer to these values than to the lower end of each range (Section \ref{sec:riss}).}  We obtain the earliest size measurements for any GRB afterglow to date, as our broad frequency coverage allows us to constrain the size even prior to the time at which DISS quenches. The RISS estimate at 8.5 days is broadly consistent with LAB18, but we find that, even for the largest angular source size allowed by our observations, the size predicted by our DISS observations is at least a factor of five times smaller than that calculated by LAB18. This may be partially due to limitations of the thin-screen approximation for the ISS modeling or to uncertainties in the LAB18 afterglow modeling, but these effects are unlikely to account for such a large discrepancy. In particular, varying afterglow parameters within the LAB18 $1\sigma$ confidence ranges changes the estimated afterglow size by only a few percent. 

The only GRB for which it has been possible to compare afterglow size estimates from ISS against a second independent observational technique is GRB\,030329, whose afterglow was resolved with VLBI at $\gtrsim 24$ days \citep{tay04,tay05,pihl07}. \cite{pihl07} note that the ISS size estimate at 15 days presented by \cite{berg03} is also smaller than an extrapolation of their VLBI observations would suggest. They propose that the discrepancy could be due to the assumed geometry of the source image. The size estimates from afterglow modeling (LAB18) given in Figure~\ref{fig:size} for GRB\,161219B and by \cite{berg03} for GRB 030329 assume that the image of the afterglow is a uniformly illuminated disk, but optically thin afterglows appear limb-brightened, meaning that the image is better modeled as a ring \citep{gran99,gl01}. This would allow DISS to persist to a larger afterglow radius, as the diffraction speckle scale would be compared to a smaller illuminated area. The correction factor is larger at both higher frequencies and later times, and may be up to a factor of $\approx 2$ for a perfect ring. If instead the GRB jet is viewed slightly off-axis and we are able to see one edge of the jet, then one side of the afterglow could be brighter due to relativistic beaming effects even prior to the nominal jet break time ($t\approx 32$ days for GRB 161219B; LAB18), creating a crescent-shaped image and a larger correction factor \citep{gdr18}. Furthermore, GRB 161219B has an unusually low radiative efficiency ($E_{\gamma}/E_K\approx4$\%; LAB18), consistent with an off-axis geometry in which the energy in the prompt emission appears low due to being relativistically beamed away from the observer. \cite{ryan15} suggest that such off-axis viewing angles may be common in GRBs. In this case, the size inferred from ISS would be smaller than the LAB18 model prediction.

The LAB18 model predicts that GRB 161219B's afterglow emission is dominated by the RS at 1.5 and 4 days at all radio frequencies, with the FS beginning to contribute at 8.5 days. The synchrotron self-absorption frequency of the RS is $\gtrsim 8$ GHz at 1.5 days, so the afterglow should be minimally limb-brightened and our first size estimate should be minimally affected for a perfectly on-axis source.  At 4 days, the afterglow is in the optically thin regime and the limb-brightening effect will be largest, while at 8.5 days the contribution of the FS emission should decrease this effect somewhat. Geometric effects are thus a plausible explanation for the changing ratio between our ISS size estimates and the LAB18 model at 4 days and 8.5 days, but given the strong LAB18 preference for a high RS self-absorption frequency at 1.5 days we require strong beaming from an off-axis viewing angle or a different explanation for the size discrepancy at this epoch. 

One alternative possibility is that we are seeing evidence of substructure in the jet, which is not predicted by the standard fireball afterglow model but has been proposed to explain the highly variable GRB prompt emission and early afterglow (e.g. \citealt{sd95,lb03,lnp09,nk09,bd16}). If confirmed by ISS observations of future GRB afterglows, similar apparent size discrepancies may therefore provide a novel way to constrain the observer viewing angle and the evolution of the jet Lorentz factor, or to suggest that an update to the basic theory is needed. %We are carrying out numerical simulations to more completely understand the effects of non-axisymmetric images on ISS (Johnson et al. in prep), but a detailed discussion is beyond the scope of this work.

\section{Summary and Conclusions}
\label{sec:conc}

We present detailed radio observations of GRB\,161219B that reveal rapid spectral and temporal variability. We demonstrate that this variability is consistent with a combination of diffractive and refractive ISS. We are able to probe the strong scattering regime due to an unusually large scattering measure $SM_{-3.5}\approx {3-20}$ along the line of sight to this burst, which shifts the transition frequency between strong and weak scattering up to $\nu_{\rm ss} \approx 22$ GHz. The scattering measure is a factor of $\approx {4-25}$ higher than predicted by the NE2001 model, illustrating that the distribution of ionized material in the ISM is poorly constrained away from the Galactic plane. Our detailed observations exemplify the power of compact extragalactic sources to improve future Galactic electron density models. 

ISS also allows us to test models of the intrinsic emission from GRB afterglows by providing direct measurements of the afterglow size. For GRB\,161219B, we obtain the earliest size measurements of any GRB afterglow to date. We find that the source size is initially $\sim 10$ times smaller than the prediction based on FS and RS modeling presented in LAB18, but agrees with the model predictions at late times (8.5 days). The early size discrepancy may indicate a slightly off-axis observer viewing angle or significant substructure in the emission region, but longer radio observations with greater simultaneou$S$ bandwidth would be required to confirm these explanations for future events. 

In general, to obtain the best possible constraints on the intrinsic radio flux densities of GRB afterglows, ideally we will need to observe for one or more full cycles of variability, so that we can accurately determine the average SED. For DISS, this will mean observing for several hours per epoch with as wide a bandwidth as possible, especially in the first few days when DISS effects are strongest. Longer observations and broader simultaneous frequency coverage than the observations presented here will provide better constraints on the correlation bandwidth and characteristic timescales of the variability, leading to better constraints on the $SM$ and the distance to the scattering screen. To fully characterize RISS and obtain additional independent constraints on the size of the afterglow, we will need to continue observing every few days even at late times, so that the evolution of $m_{\rm ref}$ can be better constrained. In time, radio observations of a population of bright GRB afterglows can better constrain both GRB physical models and the properties of the ISM away from the Galactic plane.

\begin{acknowledgements} We thank R.~Narayan and D.~Frail for useful conversations. K.D.A.~and E.B.~acknowledge support from NSF grant AST-1714498 and NASA ADA grant NNX15AE50G. K.D.A. additionally acknowledges support provided by NASA through the NASA Hubble Fellowship grant \#HST-HF2-51403.001-A awarded by the Space Telescope Science Institute, which is operated by the Association of Universities for Research in Astronomy, Inc., for NASA, under contract NAS5-26555. T.L.~is a Jansky Fellow of the National Radio Astronomy Observatory (NRAO). W.F.~acknowledges support for Program number HST-HF2-51390.001-A, provided by NASA through a grant from the Space Telescope Science Institute, which is operated by the Association of Universities for Research in Astronomy, Incorporated, under NASA contract NAS5-26555. A.G.~acknowledges the financial support from the Slovenian Research Agency (research core funding No. P1-0031 and project grant No. J1-8136) and networking support by the COST Action GWverse CA16104. VLA observations were taken as part of our VLA Large Program 15A-235 (PI: E. Berger). The NRAO is a facility of the National Science Foundation operated under cooperative agreement by Associated Universities, Inc. %This work made use of data supplied by the UK {\it Swift} Science Data Centre at the University of Leicester. 
\end{acknowledgements}

\software{CASA \citep{casa}, pwkit \citep{pwkit}} 

\bibliographystyle{aasjournal}
\bibliography{161219B_v2}

\end{document}